\documentclass[preprint,tightenlines,showpacs,amsmath,amssymb]{revtex4}
\usepackage{epsfig,epsf,graphics,psfrag}
\usepackage{times}
\usepackage{float}
\usepackage{booktabs}
\usepackage{color}
\usepackage{amsfonts,amssymb,stmaryrd,latexsym,amsmath}
\usepackage{multirow}
\usepackage{array} 
\usepackage{mathrsfs}

\usepackage{enumerate}
\usepackage{color}
\usepackage{subfigure}
\usepackage{hhline}
\usepackage[hyperindex=true]{hyperref}

\begin{document}

\bibliographystyle{unsrt}

\title{The strong decay patterns of $Z_c$ and $Z_b$ states in the relativized quark model}

\author{Guang-Juan Wang$^{1}$}\email{wgj@pku.edu.cn}
\author{Xiao-Hai Liu$^2$}\email{xiaohai.liu@fz-juelich.de}
\author{Li Ma$^{3}$}\email{ma@hiskp.uni-bonn.de}
\author{Xiang Liu$^{4,5}$}\email{xiangliu@lzu.edu.cn}
\author{Xiao-Lin Chen$^{1}$}\email{chenxl@pku.edu.cn}
\author{Wei-Zhen Deng$^{1}$}\email{dwz@pku.edu.cn}
\author{Shi-Lin Zhu$^{1,6}$}\email{zhusl@pku.edu.cn}

\affiliation{
$^1$Department of Physics and State Key Laboratory of Nuclear Physics and Technology and Center of High Energy Physics, Peking University, Beijing 100871, China\\
$^2$ Forschungszentrum J{\"u}lich, Institute for Advanced Simulation, Institut f{\"u}r Kernphysik and J{\"u}lich Center for Hadron Physics, D-52425 J¨ulich, Germany\\
$^3$ Helmholtz-Institut f\"ur Strahlen- und Kernphysik and Bethe
Center for Theoretical Physics, \\Universit\"at Bonn,  D-53115 Bonn, Germany\\
$^4$Research Center for Hadron and CSR Physics, Lanzhou University and Institute of Modern Physics of CAS, Lanzhou 730000, China\\
$^5$School of Physical Science and Technology, Lanzhou University, Lanzhou 730000, China\\
$^6$Collaborative Innovation Center of Quantum Matter, Beijing
100871, China}

\date{\today}

\begin{abstract}

Employing the relativized quark model and the quark-interchange model,
we investigate the decay of the charged heavy quarkonium-like states $Z_c(3900)$, $Z_c(4020)$, $Z_c(4430)$, $Z_b(10610)$ and $Z_b(10650)$ into the ground and radially excited heavy quarkonia via emitting a pion meson. The $Z_c$ and $Z_b$ states are assumed to be hadronic molecules composed of open-flavor heavy mesons. The calculated decay ratios can be compared with the experimental data, which are useful in judging whether the molecule state assignment for the corresponding $Z_c$ or $Z_b$ state is reasonable or not. 
The theoretical framework constructed in this work will be helpful in revealing the underlying structures of some exotic hadrons.

\end{abstract}

\maketitle

\section{Introduction}

In the last decade, numerous exotic states were observed in experiments, which results in a renaissance of the study on hadron spectra. Among those exotic hadrons, some of them are unambiguously beyond the conventional $q\bar{q}$ or $qqq$ model, such as the charged heavy quarkonium-like states $Z_c$ and $Z_b$ \cite{Choi:2007wga,Aaij:2014jqa,Chilikin:2013tch,Chilikin:2014bkk,Ablikim:2013xfr, Ablikim:2013wzq,Ablikim:2013mio,Belle:2011aa,Adachi:2012cx,Esposito:2014hsa,Albaladejo:2015lob,Pilloni:2016obd} and the heavy pentaquark candidates $P_c(4380)$ and $P_c(4450)~$\cite{Aaij:2015tga}.%, and the recently observed state $Z_c(4100)$~\cite{Aaij:2018bla}. 
Various theoretical interpretations concerning the intrinsic structures of these exotic states have been proposed in literature, such as the threshold effect~\cite{Szczepaniak:2015eza,Swanson:2014tra,Swanson:2015bsa,Bugg:2008wu,  Guo:2015umn,Mikhasenko:2015vca,Liu:2015fea,Liu:2013vfa,Liu:2014spa,Liu:2015taa},  tetraquark state~\cite{ Faccini:2013lda,Wang:2017lot,Ebert:2008kb,Patel:2014vua, Deng:2014gqa,Deng:2015lca,Zhao:2014qva,Chen:2010ze, Qiao:2013dda,Liu:2008qx, Maiani:2007wz}, hadronic molecule state \cite{Meng:2007fu,Kang:2016ezb,Liu:2008xz, Cleven:2013sq,Wang:2013cya, Ding:2008mp,Aceti:2014uea,He:2015mja,Karliner:2015ina,Chen:2015ata,Zhang:2013aoa,Cui:2013yva,Wang:2013daa,Cui:2013vfa,Khemchandani:2013iwa,Lin:2017mtz}, hadro-quarkonium state\cite{Dubynskiy:2008mq,Alberti:2016dru,Li:2013ssa,Anwar:2018bpu}. We refer to Refs. \cite{Chen:2016qju,Guo:2017jvc,Esposito:2016noz, Ali:2017jda} for a recent review about these studies.

An intriguing characteristic of those $XYZ$ states is that most of them are located close to two-particle thresholds, which inspires many theorists to regard the $XYZ$ states with this characteristic as the candidates of hadronic molecules, i.e., bound systems of two hadrons analogous to conventional nuclei. 
What we are interested in this work is the charged heavy quarkonium-like states $Z_c(3900)$, $Z_c(4020)$, $Z_c(4430)$, $Z_b(10610)$, and $Z_b(10650)$. They stay in the vicinity of $D^*\bar{D}$, $D^*\bar{D}^{*}$, $\bar{D}D^*(2S)$ (or $\bar D^* D(2S)$), $B^*\bar{B}$ and $B^*\bar{B}^*$ threshold, respectively. Correspondingly, these $Z_c$ and $Z_b$ states $^{[1]}$\footnotetext[1]{If not stated, we use $Z_c$ ($Z_b$) to represent an arbitrary charged charmonium-like (bottomonium-like) state in the following sections. } can be regarded as the hadronic molecules composed of these open-flavor meson pairs. 
The decay patterns of $Z_c$ and $Z_b$ also show some interesting characteristics. Both the valence-quark contents and spin-parity quantum numbers of $Z_c(3900)$ and $Z_c(4020)$ are the same. As hadronic molecule candidates, they should have the similar decay patterns in the heavy quark limit \cite{Bondar:2011ev,Du:2016qcr,Voloshin:2012dk}. However, for the hidden-charm channels, the existence of $Z_c(3900)$ was only confirmed in the $J/\psi\pi$ invariant mass spectrum; The $Z_c(4020)$ was observed in the $h_c\pi$ channel and has a mild signal in the $\psi(2S)\pi$, but no obvious signal is observed in the $J/\psi\pi$ channel \cite{ Ablikim:2013mio,Ablikim:2013wzq, Ablikim:2013xfr,Xiao:2013iha,Ablikim:2014dxl}. There are structures around $4.02$ GeV and $3.9$ GeV observed in $\psi(2S)\pi$ distributions, but the current experiment conclusion is still indefinite due to the complexity of the data \cite{Ablikim:2017oaf}. Besides, it is found that another charged state
$Z_c(4430)$ prefers to decay into $\psi(2S)\pi$ instead of $J/\psi\pi$~\cite{Choi:2007wga,Aaij:2014jqa,Chilikin:2013tch,Chilikin:2014bkk}. These observations are challenging both the theoretical and experimental understanding of the intrinsic structures of exotic hadrons.

Under the molecular state ansatz, a nonrelativistic constituent quark model was introduced in Ref. \cite{Liu:2014eka} to estimate the decay amplitudes of $Z_c$ and $Z_b$, and the numerical results favored the molecular state assignments for $Z_c$ and $Z_b$ by comparing with experiments. But there are several theoretical uncertainties left in Ref. \cite{Liu:2014eka}, which may affect the numerical results significantly. For instance, it is not a good approximation to treat the pion meson, the lightest Nambu-Goldstone boson, as a nonrelativistic system. In addition, the relativistic effect of the light quark in the $Q\bar q$ system is supposed to be even larger than that in the $q\bar q$ mesons, and the wave-functions of charmed and bottom mesons obtained in the nonrelativistic quark model may not work very well. The wave-functions which reflect the long-distance behavior of hadronic molecules are also ignored in Ref.~\cite{Liu:2014eka}.
Taking into account that the scattering amplitude might be very sensitive to the potentials and some relevant spatial wave functions, in this work we attempt to use a relativized quark model to improve the results. More decay channels, such as the one involving $P$-wave heavy quarkonium, will also be studied.

The article is arranged as follows: In Sec.~\ref{sec-model}, the relativized quark model and the quark-interchange model are introduced to describe the hadronic molecule decaying into one heavy quarkonium state and one light meson. The numerical results concerning the branching fraction ratios are displayed in Sec.~III. A summary is given in Sec.~IV.

\section{The Model}\label{sec-model}
 \subsection{The relativized quark model}\label{QM}

The relativized quark model is employed in our calculation due to its success in describing both the heavy and light meson spectra \cite{Godfrey:1985xj}. 
For the quark-quark interaction $i({p}_i)j({p}_j)\rightarrow i({p'}_i)j({p'}_j)$, the explicit effective Hamiltonian in the momentum space reads
\begin{eqnarray}
\label{w3}
&&H_{I_{ij}}(q)= \frac{\lambda_i}{2}\frac{\lambda_j}{2} \sum_a V^{{ij}}_a =\frac{{\lambda}_{i}}{2}  \frac{{\lambda}_{j}}{2}\left [V_{c}(q)+V_{l}(q)+V_{h}(q)+V_{so}(q)+V_{t}(q)\right],
\end{eqnarray}
where the $p_{i(j)}$ and $p'_{i(j)}$ are the momenta of the quark $i(j)$ in the initial and final states. The ${\lambda}$ is the Gell-Mann matrix. For an antiquark, it is replaced by $-{\lambda}^*$. The $V_c$, $V_l$, $V_h$, $V_{so}$ and $V_t$ represent the one-gluon-exchange (OGE) Coulomb-like interaction, linear confinement interaction, hyperfine interaction, spin-orbit interaction, and tensor interaction, respectively. Their explicit forms are 
\begin{eqnarray}
\label{potential}
V_c(q)&=&\sum_k{\omega^{\frac{1}{2}}_{ij}}\frac{4\pi\alpha_ke^{-\frac{q^2}{4\tau^2_{kij}}}}{q^2}{\omega^{\frac{1}{2}}_{ij}},  \nonumber \\
V_l(q)&=&\frac{6\pi b}{q^4}e^{-q^2/{4\sigma_{ij}^2}}, \nonumber \\
V_h(q)&=&-\sum_k{\rho_{ij}^{1+\frac{1}{2}\epsilon_{const}}}\frac{8\pi\alpha_ke^{-\frac{q^2}{4\tau^2_{kij}}}}{3m_im_j}\mathbf{s}_i\cdot\mathbf{s}_j{\rho_{ij}^{1+\frac{1}{2}\epsilon_{const}}},\nonumber \\
V^G_{so}(q)&=&\sum_k\frac{4\pi\alpha_ke^{-\frac{\mathbf{q}^2}{4\tau^2_{kij}}}}{q^2}\Big[\rho^{1+\frac{1}{2}\epsilon_{so(v)}}_{ii}{\frac{i(\mathbf{q}\times\mathbf{P}_i)\cdot{\mathbf{s}_i}}{2m^2_i}}{\rho^{1+\frac{1}{2}\epsilon_{so(v)}}_{ii}}-\rho^{1+\frac{1}{2}\epsilon_{so(v)}}_{jj}{\frac{i(\mathbf{q}\times\mathbf{P}_j)\cdot\mathbf{s}_j}{2m^2_j}}{\rho^{1+\frac{1}{2}\epsilon_{so(v)}}_{jj}}\nonumber \\
&-&\rho^{1+\frac{1}{2}\epsilon_{so(v)}}_{ij}{\frac{i(\mathbf{q}\times\mathbf{P}_j)\cdot\mathbf{s}_i-i(\mathbf{q}\times\mathbf{P}_i)\cdot\mathbf{s}_j}{m_im_j}}{\rho^{1+\frac{1}{2}\epsilon_{so(v)}}_{ij}}\Big],\nonumber \\
V^l_{so}(q)&=&-\frac{6\pi b}{q^4}e^{-q^2/{4\sigma_{ij}^2}}\Big[\rho^{1+\frac{1}{2}\epsilon_{so(s)}}_{ii}{\frac{i(\mathbf{q}\times\mathbf{P}_i)\cdot{\mathbf{s}}_i}{2m^2_i}}{\rho^{1+\frac{1}{2}\epsilon_{so(s)}}_{ii}}-\rho^{1+\frac{1}{2}\epsilon_{so(s)}}_{jj}{\frac{i(\mathbf{q}\times\mathbf{{P}}_j)\cdot\mathbf{s}_j}{2m^2_j}}{\rho^{1+\frac{1}{2}\epsilon_{so(s)}}_{jj}}\Big],\nonumber \\
V_{t}(q)&=&{\rho_{ij}^{1+\frac{1}{2}\epsilon_{tens}}}\frac{4\pi\sum_k\alpha_ke^{-\frac{q^2}{4\tau^2_{kij}}}}{m_im_jq^2}[({\mathbf{s}}_i\cdot{\mathbf{q}})( {\mathbf{s}}_j\cdot{\mathbf{q}} )-\frac{q^2}{3} \mathbf{s}_i\cdot\mathbf{s}_j]{\rho_{ij}^{1+\frac{1}{2}\epsilon_{tens}}},
\end{eqnarray}
with $\mathbf{q}=\mathbf{p}_{i(j)}-\mathbf{p}'_{i(j)}$, and $\mathbf{P}_{i(j)}=\frac{\mathbf{p}_{i(j)}+\mathbf{p}'_{i(j)}}{2}$. The $\mathbf{s}_{i(j)}$ and $m_{i(j)}$ represent the spin operator and mass of the quark with index $i$($j$), respectively. The spin-orbit interaction is divided into two parts, i.e., $V_{so}=V^G_{so}+V^l_{so}$, where the superscripts $G$ and $l$ indicate the interactions arising from the OGE and the linear confinement potentials, respectively. 

The $\alpha(q^2)$ is the running coupling constant calculated in perturbative QCD and parametrized by three Gaussian functions as follows   to simplify the numerical calculation,
\begin{eqnarray}
\alpha(q^2)=\frac{12\pi}{(33-2N_f)ln(q^2/\Lambda^2)}\approx \sum^{3}_{k=1}\alpha_ke^{-q^2/4{\gamma}^2_k},
 %=0.25e^{-q^2}+0.15e^{-q^2/10}+0.20e^{-q^2/1000}
\end{eqnarray}
where $\Lambda$ denotes the QCD scale and $ \Lambda=200$ MeV. $N_f$ is the number of the quark flavors which satisfy $4m^2_f<q^2$ ($m_f$ is the quark mass).  $k$ denotes the index of the Gaussian function. The $\alpha_k$ is the coefficient of  the $k$th gaussian function in the parametrization, and $\gamma_{k}$  is its oscillating parameter.

Compared with the nonrelativistic quark model employed in Ref.~\cite{Liu:2014eka}, two factors
\begin{eqnarray}
\omega_{ij}=1+\frac{p_ip_j}{E_iE_j}~~\text{and}~~\rho_{ij}=\frac{m_im_j}{E_iE_j},
\end{eqnarray}
are introduced  to describe the dependence of the potentials on the momenta of the interacting quarks.  Moreover, a smearing function ${\sigma^3_{ij}\over \pi^{3/2}}e^{-\sigma^2_{ij}r^2}$ is introduced to account for the nonlocal effect, since the interactions depend on both $\mathbf{P}_{i(j)}$ and $\mathbf{q}$. The relevant parameters in Eq.~(\ref{potential}) read
\begin{eqnarray}
 \label{w42}
\sigma_{ij}&=&\sigma_0(\frac{1}{2}+\frac{1}{2}(\frac{4m_im_j}{(m_i+m_j)^2})^4)+s^2(\frac{2m_im_j}{m_i+m_j})^2,\nonumber\\
\tau^{-2}_{kij}&=&{\gamma_k}^{-2}+{\sigma_{ij}}^{-2}.
\end{eqnarray}
The values of all the parameters referred are determined by fitting the mass spectra of mesons and are listed in Table \ref{par}.

\begin{table}
\caption{The values of the parameters. }\label{par}
\begin{tabular}{cccccccccc}
\toprule[1pt]\toprule[1pt]
$m_{u(d)}$ {(}GeV{)} & $m_{c}$ {(}GeV{)} & $m_{b}$ {(}GeV{)} & $b$ {(}$\text{GeV}^{2}${)} & $\epsilon_{const}$ & $\epsilon_{so(v)}$ & $\epsilon_{so(s)}$ & $\epsilon_{tens}$ & $\sigma_{0}$ {(}GeV{)} & $s$\tabularnewline
%\midrule[1pt]
$0.220$ & $1.628$ & $4.977$ & $0.18$ & $-0.168$ & $-0.035$ & $0.055$ & $0.025$ & $1.80$ & $1.55$\tabularnewline
%\midrule[1pt]
$\alpha_{s}$ & $\alpha_{1}$ & $\alpha_{2}$ & $\alpha_{3}$ & $\gamma_{1}$ (GeV) & $\gamma_{2}$ (GeV)& $\gamma_{3}$ (GeV)& $\text{\ensuremath{\Lambda}}$ (MeV) & & \tabularnewline
%\midrule[1pt]
$0.60$ & $0.25$ & $0.15$ & $0.20$ & $0.5$ & $1.58$ & $15.81$ & 200 &  & \tabularnewline
\bottomrule[1pt]\bottomrule[1pt]
\end{tabular}
\end{table}

The relevant spectra calculated in the relativized quark model are displayed in Tables \ref{massspectrumD} and \ref{massspectrumB} in Appendix A, where the nonrelativistic quark model results and the experimental data are also listed for comparison. For the heavy quarkonium, the relativistic effects can be neglected due to the large masses of the heavy quarks. The mass spectra match the experimental data well in both the nonrelativistic and relativized quark models. However, the mass spectra of the open-flavor mesons in the  relativized quark model fit the experimental data much better than those in the nonrelativistic quark model, especially for the radially excited states. It indicates that the relativistic effects are important in the open-flavor meson regime. For the light mesons, the relativistic effects are also not negligible. In Ref. \cite{Godfrey:1985xj}, the authors showed that  the mass spectra of the light mesons and their excitations are well reproduced in the relativized quark model. Therefore one can also expect that the relevant decay amplitudes calculated in the relativized quark model are more reliable than those in the nonrelativistic quark model.

\subsection{The quark-interchange model}\label{sec2}

The exotic heavy quarkonium-like states  $Z_c(3900)$, $Z_c(4020)$, $Z_b(10610)$ and $Z_b(10650)$ are generally supposed to be the hadronic molecules composed of $\bar DD^{\ast}+c.c.$, $\bar D^{\ast}D^{\ast}$, $\bar{B}B^{\ast}+c.c.$ and $\bar{B}^{\ast}B^{\ast}$, respectively. This is mainly because their masses are close to the thresholds of the corresponding components. However, the hadronic molecule interpretations   are not well established yet. To understand these exotic states better, it is necessary to study their properties from various aspects. The strong decay modes of a hadron usually have close connections with its intrinsic structure.

For a hadronic molecule, we can describe its strong decays  in terms of the near threshold scattering between the two hadron components. 
We consider the meson-meson scattering process
\begin{eqnarray} 
A(q\bar Q)+B(Q\bar q)\rightarrow C(q\bar q) +D(Q\bar Q),
\end{eqnarray}
where $q$ ($\bar q$) and  $Q$ ($\bar Q$) are the light and heavy quarks (antiquarks) in the mesons. To calculate the amplitude at the quark level, we employ the Barnes-Swanson quark-interchange model introduced in Refs. \cite{Ackleh:1991dy,Wong:2001td,Barnes:2003dg}. 
In this model, the meson-meson scattering amplitudes are evaluated at Born order with the interquark Hamiltonian, which are decomposed as
\begin{eqnarray}\label{prior-post}
H=\sum^4_{i=1}\sqrt{{\mathbf{p}}_i ^2+m^2_i}+\sum_{i<j}H_{I_{ij}}=H^0_{AB}+H^I_{AB}=H^0_{CD}+H^I_{CD},
\end{eqnarray}
where  $H^0$ is the Hamiltonian of two free mesons, $H^I_{AB}$ ($H^I_{CD}$) represents the interactions between the mesons $A$ and $B$ ($C$ and $D$).  For a molecular state decaying into a heavy qaurkonium and a light meson, the heavy quark and antiquark in the initial open-flavor mesons have to form the final heavy quarknium state, therefore the short-range interactions are expected to play the dominant role in such decays. The molecular state wave function can account for part of the long-range effects, which will be discussed later. In Ref.~\cite{Capstick:1986bm}, the three-quark interactions in the baryons are treated perturbatively. Similarly, we do not take into account the three-quark and four-quark interactions in this work.

\begin{figure}[tb]
  \centering
  \includegraphics[width=0.5\hsize]{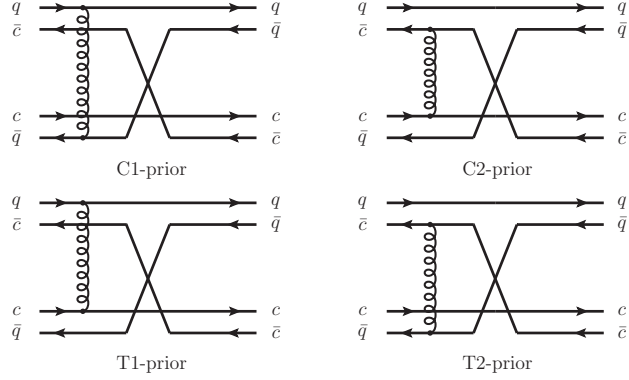}\\
  \caption{Prior diagram of scattering process $AB\rightarrow CD$. The curly line denotes the interactions between the quarks. }\label{prior}
  \end{figure}

\begin{figure}[tb]
  \centering
  \includegraphics[width=0.5\hsize]{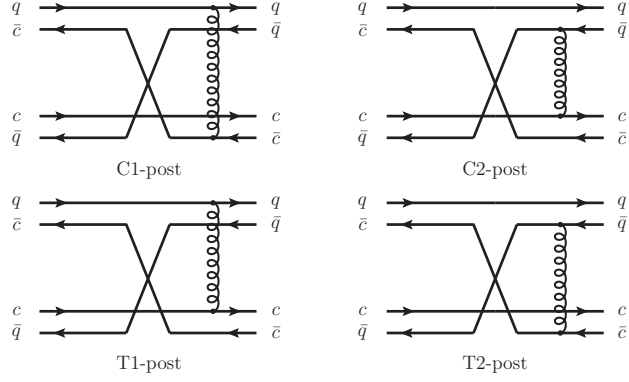}\\
  \caption{Post diagram of scattering process $AB\rightarrow CD$.}\label{post}
  \end{figure}

 According to Eq. (\ref{prior-post}), we obtain the ``Prior" and  the ``Post" $T$-matrix elements as illustrated in Fig.~\ref{prior} and Fig.~\ref{post}, respectively. Their difference is referred as the ``Prior-Post" ambiguity, which introduces the uncertainty to the decay widths and is expected to vanish if all of the pertinent wave functions are precise solutions of $H^0$ \cite{Ackleh:1991dy}. In this work, we take the average values of the ``Prior" and  ``Post"  decay widths  to calculate the decay ratios. 

At the quark level, the amplitude for a hadronic molecule $Z_c$ ($Z_b$) decaying into a charmonium (bottomonium) via emitting a light meson reads
\begin{eqnarray}
\label{w6}
T^{J}_{J_z} &=& \langle [\Psi_C(q\bar q)\Psi_D(Q \bar Q)\varphi_{CD}^{rel}(q\bar q, Q\bar 	Q)]^{J'}_{J'_z}|{\sum_{i<j} H_{I_{ij}} }|  [\Psi_A(q \bar Q) \Psi_B(Q \bar q)  \varphi_{AB}^{rel}]^J_{J_z}\rangle \nonumber\\
&=&I_{\text{flavor}}\times I_{\text{color}} \times I_{\text{spin-space}}, \nonumber \\
\Psi&=&  \phi_c\otimes \phi_f\otimes \varphi,\nonumber \\
I_{\text{favor}} &=& \langle \phi_f(C)\phi_f(D)|\phi_f(A)\phi_f(B)\rangle , \nonumber \\
I_{\text{color}}&=&\langle \phi_c(C)\phi_c(D)|\frac{\lambda_i}{2}\frac{\lambda_j}{2} |\phi_c(A)\phi_c(B)\rangle , \nonumber \\
I_{\text{spin-space}}&=&\delta_{J_zJ'_z}\langle [\varphi_C\varphi_D\varphi_{CD}^{rel}]^{J'}_{J'_z}|{\sum_{i<j}\sum_a  V^{{ij}}_a }|  [\varphi_A \varphi_B\varphi_{AB}^{rel}]^J_{J_z}\rangle, 
\end{eqnarray}
where  $J$ $(J')$ and $J_z$ $(J'_z)$ are the total angular momentum and its $z$-component of the initial (final) state.  $\Psi$ is the meson wave function obtained in the relativized quark model in Sec. \ref{QM}. It is composed of  the $\phi_f$, $\phi_c$, and $\varphi$, which represent the flavor, color, and spin-space wave functions$^{[2]}$\footnotetext[2]{In this paper we work in the momentum space to calculate the amplitude. It is of course also feasible to work in the coordinate space.}, respectively. Correspondingly, the $T$-matrix element is factored into the product of three matrix elements $I_{\text{flavor}}$, $I_\text{{color}}$, and $I_{\text{spin-space}}$.  In the flavor space, the $I_{\text{flavor}}$ cancels out when we calculate the  branching fraction ratios of the molecular states decaying into the ground and radially excited heavy quarkonium states. The $I_\text{{color}}$ takes $\frac{4}{9}$ for $qq$ and $\bar q \bar q$,  and $-\frac{4}{9}$ for $q \bar q$ interactions, respectively. 
The $\varphi_{AB}^{rel}$ ($\varphi_{CD}^{rel}$) represents the relative wave function of the $AB$ ($CD$) system in the momentum space. We assume the $Z_c$ ($Z_b$) state with $J^P=1^+$   to be an $S$-wave molecule and neglect the contributions from the higher orbital excitations. Then, a Gaussian wave function is introduced to approximately describe $\varphi_{AB}^{rel}$:
\begin{eqnarray}
\label{ww7}
&&\varphi_{AB}^{rel}(\mathbf P_A)=\frac{1}{\pi^{3/4}\beta^{3/2}}\text{Exp}[-\frac{{\mathbf P^2_A}}{2\beta^2}],\nonumber \\
&& r_0={\langle r^2 \rangle}^{\frac{1}{2}}=\sqrt{3\over 2}{1\over \beta}, ~~P_0={ \langle P^2_A \rangle}^{\frac{1}{2}}=\sqrt{3\over 2}{ \beta},
\end{eqnarray}
where $\mathbf P_A$ is the c.m. momentum of the constituent meson $A$, and $\beta$ is related to the root mean square radius  $r_0$ and momentum $P_0$ of the $Z_c$ ($Z_b$) state. The $r_0$-dependence of the branching fraction ratios is discussed in the next section.

The $I_{\text{spin-space}}$ in Eq. (\ref{w6}) is the matrix element in the spin and spatial space and reads
\begin{eqnarray}
\label{w7}
\small
I_{\text{spin-space}}&=&\Big\langle [\varphi_C\varphi_D\varphi^{rel}_{CD}]^{J'}_{J_z'}|{ H_{I_{ij}} }|  [\varphi_A \varphi_B\varphi^{rel}_{AB}]^J_{J_z}\Big\rangle \nonumber\\
&=&\sum_a\sum_{ij}\Big \langle \left[[{{(\varphi_C\chi_C)^{J_C}}(\varphi_D\chi_D)^{J_D}}]^{J_{CD}} (\varphi^{rel}_{CD})^{L_{CD}}\right]_{J'_z}^{J'} |
V^{ij}_a|\, \left [{(\varphi_A\chi_A)^{J_A}(\varphi_B\chi_B)^{J_B}}\varphi^{rel}_{AB}\right ]_{J_z}^{J} \Big \rangle \nonumber\\
&=&\sum_a\sum_{ij}{\sum_{S,L,{S',L',{L''}}}}\delta_{JJ'}\delta_{J_zJ'_z}\mathscr{W}^{S,L}_{S',L',L''} (-1)^{J+S+L''}
\left \{
\begin{array}{c c c}
   S'  &   S  &  t  \\
          L  &    L''  &  J  \\
         \end{array}
\right \} 
\nonumber\\
 & \times & \left\langle \,\left[(\Phi_C \Phi_D)^{L'} {\varphi^{rel}_{CD}}^{L_{CD}}\right]^{L''}  | |f(q^2)v^t(\mathbf  q)||\,
\Big (\Phi_A\Phi_B\Phi_{AB} \Big)^{L} \right\rangle
\nonumber\\
 & \times &
\Big\langle (\chi_C\chi_D)^{S'}|| v^t(\mathbf  s)||(\chi_A\chi_B)^{S}\Big\rangle,
\end{eqnarray}
where
\begin{eqnarray}
\mathscr{W}^{S,L}_{S',L',L''} &=&(-1)^{{L_{CD}}+J_{CD} +S'+L''}\hat{S} \hat{L} \hat{J_A} \hat{J_B}\hat{S'} \hat{L'} \hat{J_C} \hat{J_D}\hat{L''}\hat{J_{CD}} \nonumber\\
&\times & \left \{
\begin{array}{c c c}
          S_A   &    S_B   &  S  \\
          L_A   &    L_B   &  L  \\
          J_A   &    J_B   & J
      \end{array}
\right \}\left \{
\begin{array}{c c c}
          S_C   &    S_D   &  S'  \\
          L_C   &    L_D   &  L'  \\
          J_C   &    J_D   & J_{CD}
      \end{array}
\right \}\left \{
\begin{array}{c c c}
          L_{CD}   &   L'   &  L''  \\
          S'  &    J'   &  J_{CD} \\
         \end{array}
\right \},
\end{eqnarray}
with $\hat X\equiv\sqrt{2X+1}$. The $\Phi$ and $\chi$ represent the spatial and spin wave functions of pertinent mesons, respectively. For the meson $M$ ($M=A$, $B$, $C$, and $D$), $S_{M}$, $L_M$ and $J_M$ denote its spin, orbital angular momentum, and total angular momentum, respectively. 
In the $S$-wave molecular state, the $J_A$ and $J_B$ couple into the total angular momentum $J$. In the final state, the $J_C$ couples with $J_D$ to form the intermediate angular momentum $J_{CD}$. Then, the coupling between $J_{CD}$ and the relative orbital angular momentum $L_{CD}$ leads to the total angular momentum $J'$. Via the spin rearrangement, we decompose the $J$ $(J')$ into the total spin $S$ $(S')$ and the orbital angular momentum $L$ $(L'')$ of the initial (final) state with the coefficients $\mathscr{W}^{S,L}_{S',L',L''}$. The notions $|(\chi_A\chi_B)^{S} \rangle$ and $|(\chi_C\chi_D)^{S'} \rangle$ denote that the $S_{A}$ couples with $S_B$ into $S$ and $S_C$ couples with $S_D$ into $S'$, respectively. The $ |(\Phi_A\Phi_B\Phi_{AB} )^{L} \rangle$ represents that $L_A$ and $L_B$ couples into the angular momentum $L$. The notation $|\left[(\Phi_C \Phi_D)^{L'} {\varphi^{rel}_{CD}}^{L_{CD}}\right]^{L''} \rangle$ represents the coupling of the orbital angular momentums. The $L_C$ and $L_D$ couples into $L'$, $L'$ then couples with $L_{CD}$ into the total orbital angular momentum $L''$.

We also decompose the  $V^{ij}_a$ in Eq. (\ref{potential}) into the spin and momentum space by rewriting it as $V^{ij}_a=f(q^2)v^t(\mathbf s)v^t(\mathbf  q)$, where the $v^t(\mathbf s)$ ($v^t(\mathbf q)$) denote the tensor operator of order $t$ in the spin (momentum) space, and the $f(q^2)$ is the scalar part of the potential. The detailed calculations of the spin-space factor $I_\text{spin-space}$ are discussed in the following sections.

\section{Numerical results}\label{sec4}

\subsection{$S$-wave decays $Z_c\to \psi(nS)\pi$ and $Z_b\to  \Upsilon(nS)\pi$}\label{sec-numerical}

We define the branching fraction ratios as 
\begin{eqnarray}
\label{swaveratio}
R_{2}^{Z_c}=\frac{\Gamma({Z_c} \rightarrow \psi(2S) \pi)}{{\Gamma({Z_c}\rightarrow J/\psi \pi) }}, ~R_{2}^{Z_b}=\frac{\Gamma({Z_b}\rightarrow \Upsilon(2S)\pi)}{ \Gamma({Z_b} \rightarrow \Upsilon(1S) \pi) },~R_{3}^{Z_b}=\frac{\Gamma({Z_b}\rightarrow \Upsilon(3S)\pi)}{ \Gamma({Z_b} \rightarrow \Upsilon(1S) \pi) }.
\end{eqnarray}
Some of the ratios have been measured in experiments, although with large uncertainties.~We assume that the charged heavy quarkonium-like states $Z_c(3900)$,  $Z_c(4020)$ and $Z_c(4430)$ are hadronic molecules composed of $D^*\bar D$, $D^*\bar D^*$, and $\bar D D^*(2s)$ or $\bar D^* D(2s)$, respectively. To justify whether these assumptions are reasonable or not, we calculate the ratios defined in Eq. (\ref{swaveratio}) by employing the quark models introduced in Section \ref{sec-model}. 

As illustrated in Eq. (\ref{potential}) and Eq. (\ref{w7}), the spin-orbit and tensor potentials contain a  vector operator $v^{1}(\mathbf q)$ and a tensor operator $v^{2}(\mathbf q)$, respectively. They do  not contribute to the $S$-wave decays
because of $\langle L''=0||v^{1,2}(\mathbf q)||L=0\rangle=0$. The spin and spatial operators in the coulomb-like, the linear confinement and the hyperfine interactions  are scalar. Then, these potentials contribute to the S-wave decays.  Eq. (\ref{w7}) is simplified as
\begin{eqnarray}
\label{ww8}
I_\text{spin-space}=\langle\Phi_{C}\Phi_{D}{\varphi^{rel}_{CD}}|f(q^2)|\Phi_{A}\Phi_{B} {\varphi^{rel}_{AB}}\rangle\langle[\chi_{C}(q\bar q)\chi_{D}(\bar Q Q)]_{S_z}^{S}|v(\mathbf{s})|[\chi_{A}(q\bar Q)\chi_{B}(Q\bar q)]_{S_z}^{S}\rangle,
 \end{eqnarray}
where we have used the $v^{0}(\mathbf q)=1$ and omitted all the orbital angular momentums since they are $0$.  The spin operator is $v(\mathbf s)=1~\text{or} ~\mathbf{s}_i\cdot\mathbf{s}_j$.  We calculate the spin matrix elements  using  spin rearrangement and list the results in Table \ref{ss}. 
%-------------------------------------

 \begin{table}[htbp]
\caption{The matrix elements $\langle\left [\chi_C\chi_D\right]^{S}_{S_z}|{\mathbf{s}_i \cdot \mathbf{s}_j} |\,
\left [ \chi_A\chi_B \right]^{S}_{S_z} \,  \rangle$ and $\langle \left [\chi_C\chi_D\right]^{S}_{S_z} |\mathbf{1}\,|
\left [ \chi_A\chi_B \right]^{S}_{S_z} \,  \rangle$. The results of the T1 (T2) are the same in the prior and post diagrams.  The $S$ and $S_z$ denote the total spin and its $z$-component of the state. $[S_A, S_B]^S$ represents that the $S_A$ and $S_B$ combine into the total spin $S$.  }\label{ss}
\begin{tabular}{c|cccccc|c}
\toprule[1pt]\toprule[1pt]
 \multicolumn{7}{c|}{$\langle\left [\chi_C\chi_D\right]^{S}_{M}|{\mathbf{s}_i \cdot \mathbf{s}_j} |\,
\left [ \chi_A\chi_B \right]^{S}_{M} \,  \rangle$} &$\langle \left [\chi_C\chi_D\right]^{S}_{M} |\mathbf{1}\,|
\left [ \chi_A\chi_B \right]^{S}_{M} \,  \rangle$\\
 \midrule[1pt]
$[S_A,S_B]^S-[S_C,S_D]^{S}$& C1-prior  & C2-prior  &C1-post & C2-post & T1& T2 & All diagrams \\
\midrule[1pt]
$[0,1]^1-[0,1]^1$ & $-\frac{3}{8}$  & $\frac{1}{8}$ & $-\frac{3}{8}$  & $\frac{1}{8}$ &  $-\frac{1}{8}$ &$\frac{3}{8}$& $\frac{1}{2}$  \\

$[1,1]^1-[0,1]^1$ & $-\frac{3}{4\sqrt{2}}$  & $\frac{1}{4\sqrt{2}}$ & $\frac{1}{4\sqrt{2}}$ & $\frac{1}{4\sqrt{2}}$ &$-\frac{1}{4\sqrt{2}}$&$-\frac{1}{4\sqrt{2}}$ & $\frac{1}{\sqrt{2}}$ \\

$[0,1]^1-[1,1]^1$ & $\frac{1}{4\sqrt{2}}$ & $\frac{1}{4\sqrt{2}}$& $-\frac{3}{4\sqrt{2}}$ & $\frac{1}{4\sqrt{2}}$  &$-\frac{1}{4\sqrt{2}}$&$-\frac{1}{4\sqrt{2}}$  & $\frac{1}{\sqrt{2}}$ \\

$[1,1]^1-[1,1]^1$ & $0$  & $0$ & $0$ & $0$ & $-\frac{1}{2}$  &$\frac{1}{2}$  & $0$  \\
\bottomrule[1pt]\bottomrule[1pt]
\end{tabular}
%\end{center}
\end{table}

%-------------------------------------

The space factors $I_{\text{space}}\equiv\langle\Phi_{C}\Phi_{D}({{\varphi^{rel}_{CD}}})^{L_{CD}}_{m_{CD}}|f(q^2)|\Phi_{A}\Phi_{B}\varphi^{rel}_{AB}\rangle$ are the overlap integrals of the wave functions and the interaction potentials. Their explicit forms are 
 \begin{eqnarray}
 \label{w8}
I_{\text{space}}^{\text{C1-Prior}}  &=& \int d\Omega_{\mathbf{P_{C}}}\int d^{3}\mathbf{P_{A}}\int d^{3}\mathbf{p}\int d^{3}\mathbf{q} \Phi_{C}^{*}(\mathbf{q}+\mathbf{p}/2-2\mathbf{P}_{C})\Phi_{D}^{*}(\mathbf{q}-\mathbf{p}/2-\mathbf{P_{C}}-2\mathbf{P_{A}})\nonumber\\
&\times&Y^{L_{CD}*}_{m_{CD}}(\Omega_{\mathbf{P_C}}) \Phi_{A}(\mathbf{q}-\mathbf{p}/2-a \mathbf{P_{A}})\Phi_{B}(\mathbf{q}-\mathbf{p}/2-a P_{A}-2\mathbf{P_{C}}) \Phi_{AB}(\mathbf{P_{A}}) f(q^2) ,\nonumber \\
I_{\text{space}}^{\text{C2-Prior}}  &=&  \int d\Omega_{\mathbf{P_{C}}}\int d^{3}\mathbf{P_{A}}\int d^{3}\mathbf{p}\int d^{3}\mathbf{q}\Phi_{C}^{*}(\mathbf{q}+\mathbf{p}/2+\mathbf{P}_{C}-2\mathbf{P_{A}})\Phi_{D}^{*}(\mathbf{q}-\mathbf{p}/2+\mathbf{P_{C}})\nonumber \\
&\times& Y^{L_{CD}*}_{m_{CD}}(\Omega_{\mathbf{P_C}})\Phi_{A}(\mathbf{q}-\mathbf{p}/2-b \mathbf{P_{A}})\Phi_{B}(\mathbf{q}-\mathbf{p}/2-b P_{A}+2\mathbf{P_{C}})  \Phi_{AB}(\mathbf{P_{A}})f(q^2),\nonumber \\
I_{\text{space}}^{\text{C1-Post}}  &=& \int d\Omega_{\mathbf{P_{C}}}\int d^{3}\mathbf{P_{A}}\int d^{3}\mathbf{p}\int d^{3}\mathbf{q}\Phi_{C}^{*}(\mathbf{q}+\mathbf{p}/2-\mathbf{P}_{C})\Phi_{D}^{*}(\mathbf{q}+\mathbf{p}/2+\mathbf{P_{C}}-2\mathbf{P_{A}})\nonumber\\
&\times& Y^{L_{CD}*}_{m_{CD}}(\Omega_{\mathbf{P_C}})\Phi_{A}~~(\mathbf{q}-\mathbf{p}/2-a \mathbf{P_{A}})\Phi_{B}(\mathbf{q}+\mathbf{p}/2-a P_{A}-2\mathbf{P_{C}})\Phi_{AB}(\mathbf{P_{A}})f(q^2),\nonumber\\
I_{\text{space}}^{\text{C2-Post}} &=& \int d\Omega_{\mathbf{P_{C}}}\int d^{3}\mathbf{P_{A}}\int d^{3}\mathbf{p}\int d^{3}\mathbf{q} \Phi_{C}^{*}(\mathbf{q}-\mathbf{p}/2+\mathbf{P}_{C}-2\mathbf{P_{A}})\Phi_{D}^{*}(\mathbf{q}-\mathbf{p}/2+\mathbf{P_{C}})\nonumber\\
&\times &Y^{L_{CD}*}_{m_{CD}}(\Omega_{\mathbf{P_C}})\Phi_{A}(\mathbf{q}-\mathbf{p}/2-b \mathbf{P_{A}})\Phi_{B}(\mathbf{q}+\mathbf{p}/2-b P_{A}+2\mathbf{P_{C}}) \Phi_{AB}(\mathbf{P_{A}})f(q^2),\nonumber\\
I_{\text{space}}^{\text{T1}} &=& \int d\Omega_{\mathbf{P_{C}}}\int d^{3}\mathbf{P_{A}}\int d^{3}\mathbf{p}\int d^{3}\mathbf{q}\Phi_{C}^{*}(\mathbf{q}+\mathbf{p}/2-\mathbf{P}_{C})\Phi_{D}^{*}(\mathbf{q}-\mathbf{p}/2-\mathbf{P_{C}}-2\mathbf{P_{A}})\nonumber\\
&\times&Y^{L_{CD}*}_{m_{CD}}(\Omega_{\mathbf{P_C}}) \Phi_{A}(\mathbf{q}-\mathbf{p}/2-a \mathbf{P_{A}})\Phi_{B}(\mathbf{q}+\mathbf{p}/2-a \mathbf{P_{A}}-2\mathbf{P_{C}})\Phi_{AB}(\mathbf{P_{A}}) f(q^2),\nonumber\\
I_{\text{space}}^{\text{T2}}  &=& \int d\Omega_{\mathbf{P_{C}}}\int d^{3}\mathbf{P_{A}}\int d^{3}\mathbf{p}\int d^{3}\mathbf{q}\Phi_{C}^{*}(\mathbf{q}-\mathbf{p}/2+\mathbf{P}_{C}-2\mathbf{P_{A}})\Phi_{D}^{*}(\mathbf{q}-\mathbf{p}/2+\mathbf{P_{C}}\mathbf{P_{A}})\nonumber\\
&\times&Y^{L_{CD}*}_{m_{CD}}(\Omega_{\mathbf{P_C}})\Phi_{A}(\mathbf{q}-\mathbf{p}/2-b\mathbf{P_{A}})\Phi_{B}(\mathbf{q}+\mathbf{p}/2-b P_{A}+2\mathbf{P_{C}}) \Phi_{AB}(\mathbf{P_{A}})f(q^2),\nonumber
\end{eqnarray}
where
 \begin{eqnarray}
 a=\frac{m_q}{m_q+m_Q}~~,~~ b=\frac{m_Q}{m_q+m_Q},
 \end{eqnarray}
the $Y_{m_{CD}}^{L_{CD}}(\Omega_{\mathbf{P_C}})$ is the spherical harmonic function, 
the $\mathbf P_A$ ($\mathbf{P_C}$) is the c.m. momentum of the meson $A$ ($C$), and $m_q $ ($m_Q$) is the light (heavy) quark mass. The integral of each diagram due to the linear confinement potential is divergent, but the singular parts exactly cancel out when summing up all of the four diagrams (``Post" or ``Prior"), which arises from the different signs of the color factors for different diagrams. More details are given in the Appendix B.

\begin{figure}[htb]
  \centering
  \includegraphics[width=1.0\hsize]{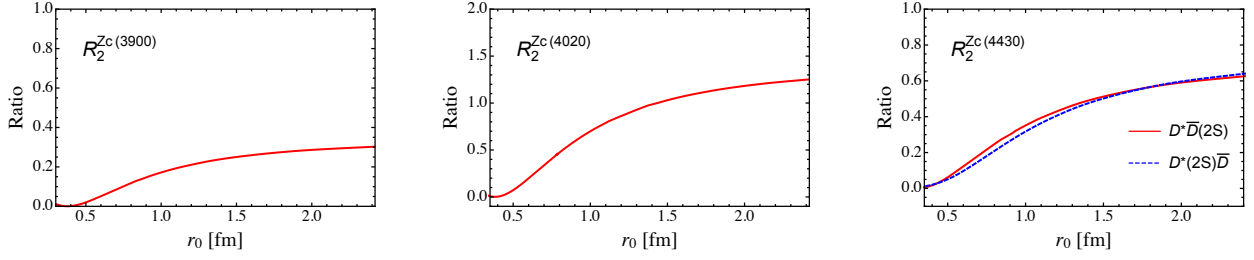}\\
  \caption{The $r_0$-dependence of the branching fraction ratios for $Z_c(3900)$, $Z_c(4020)$ and $Z_c(4430)$ decaying into $J/\psi\pi$ and $\psi(2S)\pi$.}  \label{zc}
  \end{figure}

\begin{figure}[htb]
  \centering
  \includegraphics[width=0.7\hsize]{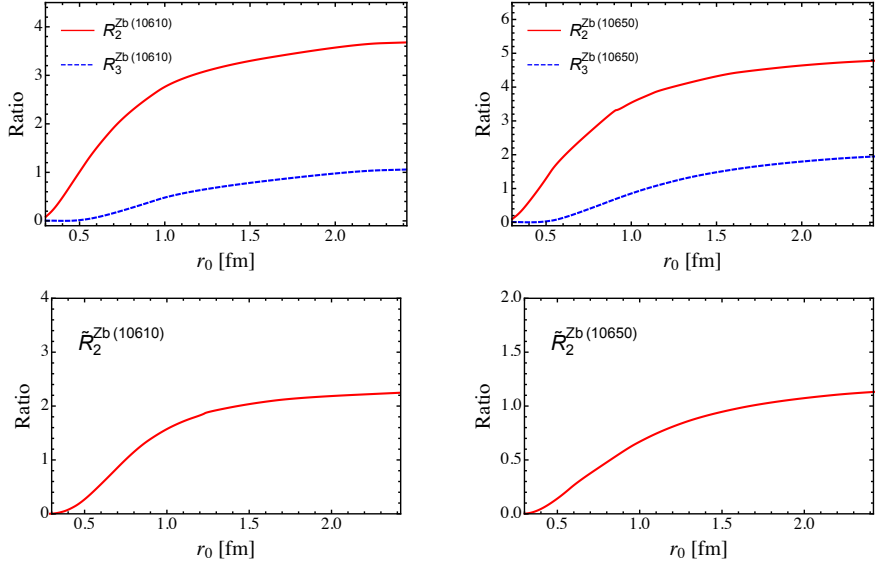}\\
  \caption{The $r_0$-dependence of the branching fraction ratios for $Z_b(10610)$ and $Z_b(10650)$ decaying into $\Upsilon(nS)\pi$, $h_b(1P)\pi$ and $h_b(2P)\pi$.}  \label{zb}
  \end{figure}

The $ r_{0}$-dependence of the branching fraction ratios are displayed  in Fig.~\ref{zc} and Fig.~\ref{zb}. It is obvious that the ratios increase with larger $ r_{0}$, which corresponds to the broader molecular wave functions.~The wave functions of the states with the radial quantum number $n$ contain $n-1$ nodes. The interaction potentials also contain nodes. When $ r_0$ is small enough, the nodes are located outside the integration. Then, the exotic state prefers to decaying into the ground heavy quarkonium via emitting a light meson due to the phase space. The decay ratio is smaller than $1$. When the $r_0$ increases, the nodes from the potential and the radial excited states may be contained in the integration. In the decay into the ground heavy quarkonium,  the parts of the integrals before and after the potential node interfere with each other destructively. In the decay into the radial excited heavy quarkonium, the nodes in the wave functions interfere with those in the potentials. This may lead to the enhancement of the decay amplitude. Thus, even with smaller phase space,  an exotic state may decay into a radial excited heavy quarkonium more easily. More interference effects are included with broader molecular wave functions.  Then, the ratio increases with larger $r_0$. When the $r_0$ is large enough, the tails of wave functions enter the integration and  slightly influence the numerical results. The decay ratios tends to be stable.

The formation of the hadronic molecules is usually supposed to be dominated by the long-range interactions between the components, for intance, the one-pion exchange potential.  For a shallow bound hadronic molecule (with mass $M$) composed of  particles A and B, the $r_0$ is estimated to be %, the size of the molecule $r_{0}$ is given by 
\begin{eqnarray}
r_0=\sqrt{\frac{1}{2\mu E_B}},
\end{eqnarray}
where $\mu=\frac{m_{A}m_{B}}{m_{A}+m_{B}}$  is the reduced mass of the constituent hadrons and $E_{B}=m_A+m_B-M$ is the binding energy of the molecule.  For the $Z_c(3900)$, $Z_c(4020)$, $Z_c(4475)$, $Z_b(10610)$ and $Z_b(10650)$ states which are located above the corresponding  thresholds, we still use the equation  to estimate their sizes with $E_B$ defined as $|m_A+m_B-M|$. The results are listed in Table \ref{size}. With these values of $r_0$, we calculate the S-wave decay ratios and list them in Table \ref{rel}.

\begin{table}
\caption{The sizes of the molecular states with the central values of the masses used in the estimation.}\label{size}
\begin{tabular}{ccccccc}
\toprule[1pt]\toprule[1pt]
 & $Z_{c}(3900)$ & $Z_{c}(4020)$ & $Z_{c}(4430)(D^{*}\bar{D}(2S))$ & $Z_{c}(4430)(D^{*}(2S)\bar{D})$ & $Z_{b}(10610)$ & $Z_{b}(10650)$\tabularnewline

$r_{0}$ {[}fm{]} & $0.9$  & $1.7$ & $0.5$ & $3$ & $1.6$ & $1.6$\tabularnewline
\bottomrule[1pt]\bottomrule[1pt]
\end{tabular}
\end{table}

\begin{table}
\caption{The S-wave decay ratios when we use the $r_0$ listed in Table \ref{size}. The experiment data is from the Refs.[71, 72]. The $R^{Z_c(4430)}$ and $R^{Z_c(4430)}_{2}$  represents the decay ratios of the $Z_c(4430)$ composed of $D^*\bar D(2S)$ and $D^*(2S)\bar D$, respectively. ``..." denotes that the corresponding experimental result is absent. }\label{rel}
\begin{center}
\begin{tabular}{ccccccccc}
\toprule[1pt]\toprule[1pt]
% & $R_{2}^{Z_{c}(3900)}$ & $R_{2}^{Z_{c}(4020)}$ & $R^{Z_{c}(4430)}(D^{*}\bar{D}(2S))$ & $R^{Z_{c}(4430)}(D^{*}(2S)\bar{D})$ & $R_{2}^{Z_{b}(10610)}$ & $R_{3}^{Z_{b}(10610)}$ & $R_{2}^{Z_{b}(10650)}$ & $R_{3}^{Z_{b}(10650)}$\tabularnewline
& $R_{2}^{Z_{c}(3900)}$ & $R_{2}^{Z_{c}(4020)}$ & $R^{Z_{c}(4430)}$ & $R_2^{Z_{c}(4430)}$ & $R_{2}^{Z_{b}(10610)}$ & $R_{3}^{Z_{b}(10610)}$ & $R_{2}^{Z_{b}(10650)}$ & $R_{3}^{Z_{b}(10650)}$\tabularnewline
Theory & $0.2$ & $1.1$ & $0.1$ & $0.7$ & $3.4$ & $0.8$ & $4.4$ & $1.6$\tabularnewline

Experiment  & $...$ & $...$ & {$\sim10$} & {$\sim10$} & $6.75\pm2.56$ & $4.00\pm1.67$ & $8.12\pm4.20$ & $9.53\pm4.80$\tabularnewline
\bottomrule[1pt]\bottomrule[1pt]
\end{tabular}
\end{center}
\end{table}

The $R_2^{Z_c(3900)}$ is much smaller than 1, indicating that the branching  fraction  of $Z_c(3900)$  into $J/\psi\pi$ is much larger than that of $\psi(2S)\pi$. 
Interestingly, $R_2^{Z_c(4020)}$ is around 1.   When $r_{0}=1.5$ fm, we find that $|T(Z_c(3900)\to \psi(2S) \pi)/T(Z_c(3900)\to J/\psi \pi)|\sim 1.8$ and $|T ( Z_c(4020)\to \psi(2S) \pi)/T( Z_c(4020)\to J/\psi \pi)|\sim 2.5$.  It implies that both the $D^*\Bar{D}$ and $D^*\Bar{D}^*$ molecules couple  to $\psi(2S)\pi$ more strongly than to $J/\psi \pi$. The smaller partial width $\Gamma(Z_c(3900)\to \psi(2S)\pi)$ is due to the fact that  the phase space of this channel is smaller, and the partial width is sensitive to the final state momentum. The $Z_c(3900)$ is observed in the $J/\psi\pi$ invariant mass spectrum, which is consistent with our prediction that the ratio $R_2^{Z_c(3900)}$ is much smaller than 1.

In the $e^+e^-\to \psi(2S)\pi^+\pi^- $ process, an obvious resonance-like structure around $4.03$ GeV is observed in the $\psi(2S)\pi^\pm$ invariant mass spectrum for data at the c.m. energy $\sqrt{s}=4.416$ GeV \cite{Ablikim:2017oaf}. This structure can be identified as the $Z_c(4020)$. The resonance-like structure around $3.9$ GeV can also be seen in $\psi(2S)\pi^\pm$ distributions at some c.m. energies, but this structure could also arise from the reflection effect of the other structure around $4.03$ GeV in the Dalitz plot. Due to the complexities of the Dalitz plots for the $e^+e^-\to \psi(2S)\pi^+\pi^- $ process at different c.m. energies,  
the BESIII collaboration did not give a definite conclusion in their paper and claimed that their fit cannot describe the data well \cite{Ablikim:2017oaf}. The experimental ratios $R_2^{Z_c(3900)}$ and $R_2^{Z_c(4020)}$ are thus still unknown.

%--------------------------
%\begin{table}
%\caption{The S-wave decay ratios with the $r_0$ varying in the molecular scale $1 \sim 2$ fm.  All the other parameters are already fixed by the mass spectra of the mesons in the quark model. The experiment data are from the Refs. \cite{shencp:2014,Garmash:2015rfd}. The $R^{Z_c(4430)}_{2}$  represents the decay ratios of the $Z_c(4430)$ composed of $D^*\bar D(2S)$ or $D^*(2S)\bar D$. ``..." denotes that the corresponding experimental result is absent. }
%\label{numericalresults}
%\begin{tabular}{ccccccccc}
%\toprule[1pt]\toprule[1pt]
% & $R_{2}^{Z_c(3900)}$ & $R_{2}^{Z_c(4020)}$  &$R_2^{Z_c(4430)}$  & $R_{2}^{Z_b(10610)}$ & $R_{3}^{Z_b(10610)}$ & $R_{2}^{Z_b(10650)}$ & $R_{3}^{Z_c(10650)}$\\
%Theory   & $0.2\sim0.3$ & $0.7\sim 1.2$  &$0.3\sim 0.6$& $2.7\sim 3.6$ & $0.4\sim1.0$ & $3.6 \sim 4.7$ & $0.9 \sim 1.8$\\
%Experiment data  & ... &  ... & $\sim 10$& $6.75\pm2.56$ & $4.00\pm1.67$ & $8.12\pm4.20$ & $9.53\pm4.80$\\
%\bottomrule[1pt]\bottomrule[1pt]
%\end{tabular}
%\end{table}
%--------------------------

The mass of $Z_c(4430)$ is close to the threshold of $\bar DD^*(2S)$ or $\bar D(2S)D^*$, and the more favorable quantum numbers are $J^P=1^+$. Due to these properties,  the $Z_c(4430)$ has ever been identified as a molecular state composed of $\bar DD^*(2S)$ or $\bar D(2S)D^*$. We display its strong decay ratios with different  $r_0$ in two configurations in Fig. \ref{zc}.
 We find the decay ratio is smaller than $1$, which is much smaller than the estimated ratio $\sim 10$ in experiments.
Without introducing any other dynamic mechanisms, this result implies that the assignment of a pure $\bar DD^*(2S)$ or $\bar D(2S)D^*$ hadronic molecule for $Z_c(4430)$ is not favourable. The ratio $R_2^{Z_c(4430)}$ calculated in this paper is different from that estimated in the naive nonrelativistic quark model \cite{Liu:2014eka}, which shows the model sensitivity of numerical results. This model sensitivity can be partly ascribed to the uncertainties of the relevant wave functions.
As listed in Tables \ref{massspectrumD} and \ref{massspectrumB}, the relativized quark model reproduces the charmed and bottomed meson spectra much better than the nonrelativistic model. Thus, the relativized quark model is more suitable in describing the hadronic molecule decays discussed in this paper.

We list the theoretical values of $R_{2,3}^{Z_b}$ in Table \ref{rel}. The calculated ratios  $R_2^{Z_b(10610)}$ and $R_2^{Z_b(10650)}$ approximately fall within the ranges of experimental values, but the theoretical ratios $R_3^{Z_b(10610)}$ and $R_3^{Z_b(10650)}$ significantly deviate from the experimental central values. However, one should also notice that the uncertainties of the experimental data are still quite large, and
the estimated ratios $R_3^{Z_b(10610)}$ and $R_3^{Z_b(10650)}$ are still of the same order as the experimental values.
As a relatively weak argument, these theoretical results to some extent can support the assumptions of identifying $Z_b(10610)$ and $Z_b(10650)$ as the $B^*\Bar{B}$ and $B^*\Bar{B}^*$ molecules, respectively.

\subsection{$P$-wave decays $Z_b \to h_b(nP) \pi$}\label{sec4b}
For the decays $Z_{c(b)}\to h_{c(b)}(nP)\pi$, there is a P-wave orbital excitation between the two hadrons  in the final state. Since the masses of $Z_c(3900)$ and $Z_c(4020)$ are supposed to be below the $h_c(2P)\pi$ threshold, we do not discuss the ratios in relevant with the $Z_c$ states. For the two $Z_b$ states, the $h_b(1P)\pi$ and $h_b(2P)\pi$, we define the branching fraction ratio
\begin{eqnarray}
\label{pwaveratio}
\tilde{R}_2^{Z_b}=\frac{\Gamma({Z_b} \rightarrow h_b(2P) \pi)}{{\Gamma({Z_b}\rightarrow h_b(1P)\pi })}.
\end{eqnarray}
In the decay process, the total spin $S=1$ in the initial state flips into the total spin $S'=0$ in the final state, while the initial orbital momentum $L=0$ flips into  $L''=1$ in the final state. Since $\langle 1||v^{0,2}(\mathbf{q}^2)||0\rangle=0$, the OGE Coulomb-like, the linear, the hyperfine and the tensor potentials do not contribute. For the spin-orbital potential, the spin operator $v^1(\mathbf s)=\mathbf{s}_i$ is a vector. The reduced matrix element for the $\mathbf{s}_q$ is,

\begin{eqnarray}
\label{w18}
&&\left \langle \, \left [\chi_C(q\bar q)\chi_D(Q\bar Q)\right]^{S'}||\mathbf{s}_q||\,
\left [ \chi_A(q\bar Q)\chi_B(Q\bar q) \right]^{S} \right\rangle\nonumber\\
&&=\sum_{S_{14},S_{23}}(-1)^{S_D+S_B-2s_Q-s_{\bar q}-s_{\bar Q}}\hat{S}_A\hat{S}_B\hat{S}_{14}\hat{S}_{23}\left \{
\begin{array}{c c c}
          s_q   &    s_{\bar c}   &  S_A  \\
          s_{\bar q}   &    s_c   &  S_B  \\
          S_{14}   &S_{23}   & S
      \end{array}
\right \}  \delta_{S_D,S_{23}}(-1)^{S+S_C+S_{13}-1}\hat{S}\hat{S'}\nonumber\\
&&\times \left \{
\begin{array}{c c c}
          S_{14}   &    S_{23}   &  S  \\
          S'   &    1  &  S_C  \\
\end{array}
\right \}(-1)^{S_{14}}\hat{S}_{14}\hat{S}_C\left \{\begin{array}{c c c}
          S_C  &   1   &  S_{14}  \\
          1/2   &   1/2  & 1/2  \\
\end{array}
\right \}\sqrt{s_q(s_q+1)(2s_q+1)},
\end{eqnarray}
where $s_{q}$ $(s_{\bar q})$ and $s_{Q}$ $(s_{\bar Q})$ are the spin of  light and heavy quarks (antiquarks), respectively.  $S_{14}$ and $S_{23}$ represent the spin of the two light and two heavy quarks in the initial state, respectively.
The calculations of the reduced matrix elements for the $\mathbf{s}_{\bar q}$, $\mathbf{s}_Q$ and $\mathbf{s}_{\bar Q}$ are similar.  We list the results   in Table \ref{wt3}. 
\begin{table}[htbp]
\caption{$\langle \, \left [\chi_C\chi_D\right]^{S'}||\mathbf{s}_q||\,
\left [ \chi_A\chi_B \right]^{S} \,  \rangle$ in Eq. (\ref{w18}). $S$ and $S'$ denote the total spin of the initial and final states, respectively.}\label{wt3}
\begin{center}
\begin{tabular}{ccccccc}
  \toprule[1pt]\toprule[1pt]
   $[S_A,S_B]^S-[S_C, S_D]^{S'}$  &$s_q$ & $s_{\bar Q}$ & $s_Q$ & $s_{\bar q}$  \\

  $[0,1]^1-[0,0]^0$ & $-\frac{\sqrt{3}}{4}$ & $\frac{\sqrt{3}}{4}$ &$-\frac{\sqrt{3}}{4}$ &$\frac{\sqrt{3}}{4}$  \\

 $[1,1]^1-[0,0]^0$   & $\frac{\sqrt{3}}{2\sqrt{2}}$ & $\frac{\sqrt{3}}{2\sqrt{2}}$ & $-\frac{\sqrt{3}}{2\sqrt{2}}$ &$-\frac{\sqrt{3}}{2\sqrt{2}}$ \\

 $ [0,1]^1-[1,1]^1$  & $-\frac{1}{4}$ & $\frac{1}{4}$ & $\frac{3}{4}$ &$-\frac{3}{4}$ \\

  $[1,1]^1-[1,1]^1$  & $\frac{1}{2\sqrt{2}}$ &  $\frac{1}{2\sqrt{2}}$ &-$\frac{1}{2\sqrt{2}}$ &- $\frac{1}{2\sqrt{2}}$ \\
  \bottomrule[1pt]\bottomrule[1pt]
\end{tabular}
 \end{center}
\end{table}

For the spatial reduced matrix, there is a relation
 \begin{eqnarray}
 \label{br}
C_{LL_{z};1\mu}^{L''L_{z}''} I_{\text{space}}&=&{C_{LL_{z};1\mu}^{L''L_{z}''}} \left \langle\left[(\Phi_{C}\psi_{D})^{L'}\Phi_{CD}^{L_{CD}}\right]^{L''}||f(q){v^{t}(q)}||[\Phi_{A}\Phi_{B}\Phi_{AB}]^{L}\right \rangle \nonumber \\
&=&\sqrt{2L''+1}C_{L'L'_{z},L_{CD}m_{CD}}^{L''L_{z}''}\langle\Phi_{C}(\Phi_{D})_{m_{D}}^{L_{D}}(\varphi_{CD}^{rel})_{m_{CD}}^{L_{CD}}|f(q)v(\mathbf{q})_{\mu}^{1}|\Phi_{A}\Phi_{B}\Phi_{AB}\rangle.
 \end{eqnarray}
For the decay $Z_b\rightarrow h_b \pi$, one has $L=L_z=0$, $t=1$, and $L_{CD}=L'=L''=1$. The calculation of the $ I_{\text{space}}$ is similar to Eq. (\ref{w8}).

The $ r_{0}$-dependence of the ratio $\tilde{R}_2^{Z_b}$ is illustrated in Fig. \ref{zb}. In this figure, we find that the $\tilde{R}_2^{Z_b}$ increases with larger $  r_{0} $. The $Z_b(10610)$ and $Z_b(10650)$ prefer to decaying into the $h_b(2P)\pi$ channel when the $ r_{0}$ are larger than $1.0$ fm and $1.7$ fm, respectively.
We list the numerical results when the $r_0$ is $1.6$ fm in Table \ref{numericalresultsp}. Our results is larger than those in Ref \cite{Cleven:2011gp}, and fall in the range of the experimental results. 
%--------------------------
\begin{table}
\caption{The P-wave decay ratios when the $r_0$ is $1.6$ fm.  The experimental data comes from Ref. \cite{Garmash:2015rfd}. }
\label{numericalresultsp}
%\begin{center}
\begin{tabular}{ccccccc}
\toprule[1pt]\toprule[1pt]
& $\tilde{R_2}^{Z_b(10610)}$ & $\tilde{R_2}^{Z_b(10650)}$\\
Theory   & $2.1$ & $1.0$ \\
Ref \cite{Cleven:2011gp} & $0.21$ & $0.27$ \\ 
Experiment data  & $1.43\pm0.85$ & $1.84\pm0.95$ \\
\bottomrule[1pt]\bottomrule[1pt]
\end{tabular}
\end{table}
%--------------------------

\section{Summary} \label{sec5}

In this work, we assume that the $Z_c$ and $Z_b$ states are hadronic molecules composed of open-flavor mesons. In the framework of the relativized quark model and the quark-interchange model,
we calculate the branching fraction ratios of $Z_c$ ($Z_b$) states decaying into ground and radially excited charmonia (bottomonia) via emitting a pion meson.  These ratios can be compared with the experimental data, which are useful in judging whether the molecule state assignment for the corresponding $Z_c$ or $Z_b$ state is reasonable or not. Our calculations indicate that the $Z_c(3900)$ and $Z_c(4020)$ have a larger coupling with $\psi(2S)\pi$ than $J/\psi\pi$. However, constrained by the phase space, the partial width $\Gamma(Z_c(3900)\to J/\psi\pi)$ is much larger than $\Gamma(Z_c(3900)\to \psi(2S)\pi)$, which is consistent with the current experimental observations. However, the explicit values of $R_2^{Z_c(3900)}$ and $R_2^{Z_c(4020)}$ still need to be checked by the future experiments. The value of $R_2^{Z_c(4430)}$ calculated in this relativized quark model is much smaller than the experiment estimation in Refs. \cite{Choi:2007wga,Aaij:2014jqa,Chilikin:2013tch,Chilikin:2014bkk}, which does not favor the assumption of identifying the $Z_c(4430)$ as a pure $\bar{D}D^*(2S)$ or $\bar D^* D(2S)$ molecule.  The ratios $R_2^{Z_b}$ and $R_3^{Z_b}$ are approximately consistent with the experimental estimations. Besides, the calculated $P$-wave decay ratio $\Gamma({Z_b} \to h_b(2P) \pi) / \Gamma({Z_b} \to h_b(1P) \pi)$ also approximately falls within the range of experimental values, which implies the $B^*\Bar{B}$/$B^*\Bar{B}$ molecule assignment for $Z_b(10610)$/$Z_b(10650)$ is favorable.

It should be stressed that our calculations are based on the assumption that the $Z_c$ and $Z_b$ states are hadronic molecules, and we use the Gaussian distribution functions to describe their relative wave functions. This simple assumption about the formalism of molecular wave functions will definitely bring some uncertainties to the numerical results. Fortunately, we notice that the decay ratios are not very sensitive to the free parameter $r_{0} $ of the wave functions.   

The theoretical framework used in this work will be helpful in revealing the underlying structures of some exotic states. And it is also very promising that the predictions based on this framework could be checked in the near future with the huge data samples accumulated by the BESIII, LHCb, Belle and Belle-II collaborations.

\subsection*{Acknowledgments}
We are grateful to the helpful discussions with Yan-Rui Liu, Jia-Jun Wu, and Yuan Song. We also thank Prof. Ulf-G. Mei{\ss}ner for a careful reading and helpful suggestions. This work is supported by the National
Natural Science Foundation of China (NSFC) under Grants No.11575008 and No.
11621131001, by the National Key Basic Research Program of
China (2015CB856700), by the NSFC and Deutsche Forschungsgemeinschaft (DFG) through 
funds provided to the Sino--German Collaborative Research Center ``Symmetries and the Emergence of Structure in QCD'' (NSFC Grant No.~11621131001,
DFG Grant No.~TRR110).

\subsection*{Appendix}

\subsection{The mass spectra}

In the relativized quark model, the kinematic term is replaced by the relativistic term $E_i=\sqrt{m_i^2+\mathbf{p}_i^2}$. We calculate the mass spectra of the heavy mesons and the heavy qaurkonia. The mass spectra of the mesons involved in this work are listed in Tables \ref{massspectrumD} and \ref{massspectrumB}.  
%------------------------
\begin{table}
%\footnotesize
\caption{Mass spectra of the charmed mesons. $M^R_{th}$, $M^{NR}_{th}$, and $M_{exp}$ are the mass spectra in the relativized quark model, the nonrelativistic quark model \cite{Liu:2014eka}, and in experiments \cite{Agashe:2014kda}, respectively. }\label{massspectrumD}
%\begin{center}
\begin{tabular}{ccccccccccccccc}
  \toprule[1pt]\toprule[1pt]
& $D$ & $D^*$ &$D(2S)$ & $D^*(2S)$ & $J/\psi$ & $\psi(2S)$  & $h_c(1P)$ & $h_c(2P)$ &$\chi_{c0}(1P)$ &$\chi_{c1}(1P)$ &$\chi_{c2}(1P)$ \\
  $M^{R}_{th}$ [GeV] & 1.873 &  2.038 & 2.582 & 2.645 & 3.091 & 3.679 &3.515 &3.956  &3.443 &3.508 &3.548  \\
$M^{NR}_{th}$ [GeV]& 1.920 & 1.993 & 2.711 & 2.769 & 3.089 & 3.701  &--&--  &-- &-- &-- \\
$M_{exp}$ [GeV] & 1.865 & 2.010 & 2.539 & 2.612 & 3.097 & 3.686  &3.525 &-- &3.414 &3.511 &3.556 \\
\bottomrule[1pt]\bottomrule[1pt]
\end{tabular}
%\end{center}
\end{table}
%------------------------

%------------------------
\begin{table}[htbp]
%\footnotesize
\caption{Mass spectra of the bottom mesons. $M^R_{th}$, $M^{NR}_{th}$, and $M_{exp}$ are the mass spectra in the relativized quark model, the nonrelativistic quark model \cite{Liu:2014eka}, and in experiments \cite{Agashe:2014kda}, respectively.} \label{massspectrumB}
%\begin{center}
\begin{tabular}{cccccccccccccc}
  \toprule[1pt]\toprule[1pt]
   & $B$  & $B^*$ &$B_1$  &$B^{\ast}_1$  &$\Upsilon(1S)$  & $\Upsilon(2S)$  & $\Upsilon(3S)$ &$h_b(1P)$ &$h_b(2P)$ &$\chi_{b0}(1P)$ &$\chi_{b1}(2P)$ &$\chi_{b2}(1P)$ \\

  $M^R_{th}$ [GeV] &5.310&5.369&5.905&5.934&9.466&10.010&10.359  &9.881 &10.251  &9.847  &9.876   &9.896 \\

  $M^{NR}_{th}$ [GeV]& 5.387 & 5.411 & 5.748 &-& 9.471 &9.944 &10.347 &--   &--  &-- &--   &-- \\

  $M_{exp}$ [GeV] & 5.279 & 5.325 &- &-& 9.460  & 10.023 & 10.355  &9.899&10.260&9.859&9.893&9.912 \\
\bottomrule[1pt]\bottomrule[1pt]
\end{tabular}
%\end{center}
\end{table}
%-------------------------

\subsection{$I_{space}$}
The linear confinement effect $V_{l}$ contributes to the $S$-wave decay amplitudes. In  Eq. (\ref{w8}), the $I_{\text{space}}$ in relevant with $V_l$ is
\begin{eqnarray}
\label{w14}
&&\int d^3\mathbf{q}e^{-\frac{u}{2}(q-q_0)^2}V_{l}=6\pi b\int d^3\mathbf{q} e^{-\frac{u}{2}(q-q_0)^2}\frac{e^{-\frac{q^2}{4\sigma^2_{ij}}}}{q^4}\nonumber\\
&&~~~~~~=-6\pi b(2\pi)^{3/2}\sqrt{z}{e^{-\frac{\mu q^2_0}{2}} }_1F_1(-\frac{1}{2},\frac{3}{2};\frac{\mu^2 q^2_0}{2z})+6{\pi}b(2\pi)e^{-\frac{\mu q^2_0}{2}}e^{-\frac{zq^2}{2}}\frac{2}{q}|_{q\rightarrow0}.\nonumber
\end{eqnarray}
where  $z=\mu+\frac{1}{2{\sigma_{ij}}^2}$. $q_0$ and $\mu$ are parameters in relevant with the momenta and masses of the mesons in the initial and final states. Their explicit forms are referred to Ref. \cite{Wong:2001td}. When $q=0$, there is $q_0=0$. The divergent terms in the Prior or Post diagrams cancel out exactly due to the color factors.

For the P-wave decays, the spin-orbital effect  $V^{G,l}_{so}$ contribute and is factorized as $f(q)\frac{(\mathbf{q}\times\mathbf{P_i})\cdot \mathbf{s}_i}{m^2}$. The $I_{\text{space}}$ is, 

\begin{eqnarray}
\label{w19}
&&I_{\text{space}}\sim \int d\mathbf{q}e^{{-\frac{\mu}{2}(\mathbf{q}-\mathbf{q}_0)^2}}f(q){q}_{\mu}\nonumber\\
&&=\frac{1}{\mu}\frac{\partial}{\partial q^{\mu}_0}\int d\mathbf{q}
e^{ -\frac{\mu}{2} (\mathbf{q}-\mathbf{q}_0)^2}f(q)+{q}_{0\mu}\int d{\mathbf{q}}e^{ -\frac{\mu}{2} (\mathbf{q}-\mathbf{q}_0)^2}f(q).\nonumber
\end{eqnarray}
The divergences arising from the  two integrals are
\begin{eqnarray}
\label{w20}
&&\frac{1}{\mu}\frac{\partial}{\partial {q}^{\mu}_0}e^{-\frac{\mu q^2_0}{2}}[\frac{\sinh(\mu qq_0)}{\mu q_0q^2}+\frac{\cosh(\mu q_0q)}{q}]|_{q\rightarrow0}+{q}_{0{\mu}}e^{-\frac{\mu q^2_0}{2} }[\frac{\sinh(\mu qq_0)}{\mu q_0q^2}+\frac{\cosh(\mu q_0q)}{q}]|_{q\rightarrow0}\nonumber\\
&&=\frac{1}{\mu}e^{-\frac{\mu q^2_0}{2} }[\frac{\cosh(\mu qq_0)}{ q^2_0q}+\frac{\mu \sinh(\mu q_0q)}{q_0}-\frac{\sinh(\mu q_0q)}{\mu q^3_0q^2}]|_{q\rightarrow0}.\nonumber
\end{eqnarray}
At $q=0$,  the two singular parts cancel out.

\end{document}